\documentclass[proc]{edpsmath}
\usepackage{float}
\usepackage{subcaption} 
\usepackage{graphicx}                  
\usepackage{caption}
\usepackage{dsfont}
\usepackage[square, numbers]{natbib}

\newcommand{\X}{\mathbf{x}}
\newcommand{\I}{\mathcal{I}}
\newcommand{\slgp}{\mathfrak{p}}
\newcommand{\densf}[1]{p_{#1}}
\newcommand{\distf}[1]{\mu_{#1}}

\newcommand{\gpmean}{m}
\usepackage{tikz}
\usetikzlibrary{positioning, fit, calc, shapes, arrows, matrix, shapes.misc}
\tikzset{cross/.style={cross out, draw=black, fill=none, minimum size=2*(#1-\pgflinewidth), inner sep=0pt, outer sep=0pt}, cross/.default={2pt}}

\begin{document}
	
	\title{Goal-oriented adaptive sampling under random field modelling of response probability distributions} 
	\author{A. Gautier}\address{Institute of Mathematical Statistics and Actuarial Science, University of Bern, Switzerland}
	\author{D. Ginsbourger}\sameaddress{1}
	\author{G. Pirot}\address{Centre for Exploration Targeting, The University of Western Australia, Australia}
	%
	%
	\begin{abstract}
		In the study of natural and artificial complex systems, responses that are not completely determined by the considered decision variables are commonly modelled probabilistically, resulting in response distributions varying across decision space. We consider cases where the spatial variation of these response distributions does not only concern their mean and/or variance but also other features including for instance shape or uni-modality versus multi-modality. Our contributions build upon a non-parametric Bayesian approach to modelling the thereby induced fields of probability distributions, and in particular to a spatial extension of the logistic Gaussian model. 
		The considered models deliver probabilistic predictions of response distributions at candidate points, allowing for instance to perform (approximate) posterior simulations of probability density functions, to jointly predict multiple moments and other functionals of target distributions, as well as to quantify the impact of collecting new samples on the state of knowledge of the distribution field of interest. In particular, we introduce adaptive sampling strategies leveraging the potential of the considered random distribution field models to guide system evaluations in a goal-oriented way, with a view towards parsimoniously addressing calibration and related problems from non-linear (stochastic) inversion and global optimisation. 
	\end{abstract}

	\maketitle
	\section{Introduction}
	
	\quad 
	Many problems in science and engineering boil down to studying the effect on a response of interest of varying some \textit{decision} or \textit{control} variables $\X$. Yet it is typically unrealistic to assume a deterministic relationship between $\X$ and the response, be it for instance because of uncertainty in other input variables or because the assumed response and/or observation generating processes themselves involve some randomness. Addressing optimisation and related tasks in such frameworks comes with both conceptual and implementation challenges. While a number of contributions around stochastic and ``noisy'' optimisation have been developed, they often rely on homoscedasticity or specific distributional assumptions. In contrast, expensive-to-evaluate systems for which various features of response distributions evolve in decision space call for versatile modelling approaches.        
	
	\quad Here we consider the situation where responses follow probability distributions $\distf \X $ indexed by decision variables $\X \in D$ and with a common support $\mathcal I$. Without loss of generality, the index set --or \textit{decision space}-- $D$ is assumed to be a compact subset of $\mathbb{R}^d$ ($d=1$ or $2$ in forthcoming examples). 
	We furthermore assume that an objective function $g(\X) = \rho(\distf \X )$ depending on $\X$ through $\distf \X $ is to be minimised, where $\rho$ returns for any considered probability distribution a real-valued quantity such as a moment or a quantile with some given level. 
	One of the main challenges in the considered framework is that typically, $g(\X)$ cannot be exactly evaluated as one only observes draws from $\distf \X $. We use the letter $t$ (resp. $T$ in random form) to denote such draws.
	
	\begin{figure}[H]
		\begin{tikzpicture}[thick,scale=1]
			\coordinate (a1) at (-0.5,0);
			\coordinate (a2) at (5,0);
			\coordinate (a3) at (8,2.25);
			\coordinate (a4) at (2.5,2.25);
			
			\filldraw[fill=gray!20] (a1) -- (a2) -- (a3) -- (a4) -- cycle node [xshift=2cm,yshift=0.25cm] {Index space $D$};
			
			\coordinate (b1) at (1.5,1);
			\coordinate (b2) at (4,0.5);
			\coordinate (b3) at (6.5,1.65);
			
			\filldraw (b1) node [below] {$\mathbf x_1$} circle (1pt);
			\filldraw (b2) node [below] {$\mathbf x_2$} circle (1pt);    
			\filldraw (b3) node [below] {$\mathbf x_3$} circle (1pt);
			
			\filldraw[fill=white, shift={(b1)}] (-1.125, 0.5) -- (-0.375, 0.5) -- (0, 0)  -- (0.375, 0.5) -- (+1.125, 0.5) -- (+1.125, 2.6) -- (-1.125, 2.6)   -- cycle node[xshift=-0.45cm,yshift=-0.025cm, text width=2.25cm] {\centering \tiny 150 data points available\\
				\begin{center}
					\includegraphics[width=0.85\textwidth]{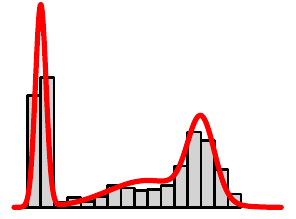}
				\end{center}
			} ;
			
			\filldraw[fill=white, shift={(b2)}] (-1.125, 0.5) -- (-0.375, 0.5) -- (0, 0)  -- (0.375, 0.5) -- (+1.125, 0.5) -- (+1.125, 2.6) -- (-1.125, 2.6)   -- cycle node[xshift=-2.95cm,yshift=0.45cm, text width=2.25cm] {\centering \tiny No data available
				\\
				\hspace{1cm} \\
				\begin{center}
					\includegraphics[width=0.85\textwidth]{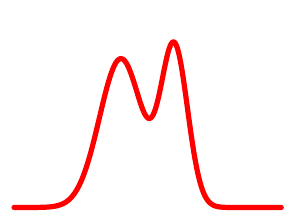}
				\end{center}
			};
			
			\filldraw[fill=white, shift={(b3)}] (-1.125, 0.5) -- (-0.375, 0.5) -- (0, 0)  -- (0.375, 0.5) -- (+1.125, 0.5) -- (+1.125, 2.6) -- (-1.125, 2.6)   -- cycle node[xshift=-5.45cm,yshift=-0.65cm, text width=2.25cm] {\centering \tiny 10 data points available\\
				\begin{center}
					\includegraphics[width=0.85\textwidth]{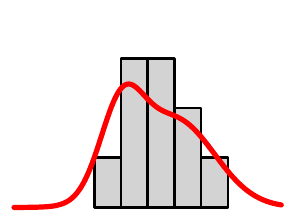}
				\end{center}
			} ;
		\end{tikzpicture}
		\caption{Schematic representation of a field of probability distributions $\distf \X $ (represented by their probability density functions, in red) to be predicted based on scattered samples (with associated histograms in grey). 
			Here the end goal is to collect additional samples towards the minimisation of $g(x)=\rho(\distf \X )$.
		}
	\end{figure}
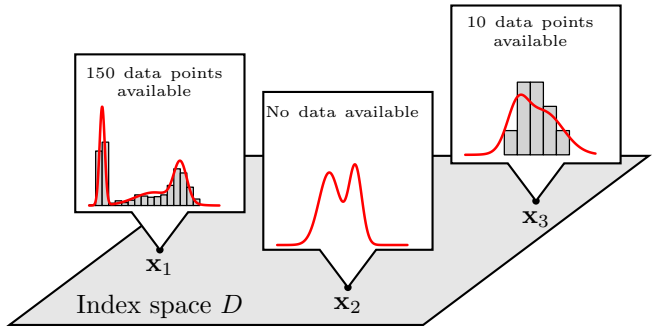
	
	\quad In stochastic optimisation (\citet{Ruszczynski2003}; \citet{Birge2011}; \citet{Prekopa2013}), having adapted assumptions for $\distf \X $ is important to achieve good results.
	Approximation procedures have been studied, including but not limited to the Robbins Monro procedures and further developments in Bayesian Approximation (\citet{Robbins1951}; \citet{Mandt2017}) or the multi-armed bandit paradigm (\citet{Thompson1933}; \citet{Cesa2006}; \citet{Bubeck2012}). The most natural choice for the functional $\rho$ is to consider the  expectation. 
	Yet, many other choices for $\rho$ have been considered. These choices include quantiles (\citet{Rostek2010}; \citet{Torossian2020}), the conditional value at risk (\citet{Rockafellar2000}) or expectiles (\citet{Bellini2015}). 
	It is also possible to learn the unknown distribution $\distf \X $ (\citet{Hall2004}; \citet{Efromovich2010}; \citet{Moutoussamy2015}; \citet{zhu2020replication}) but it generally requires a large sample size.
	
	\quad 
	In the context of global optimisation under expensive objective function evaluations, Bayesian Optimisation (BO) leverages the  potential of meta-modelling, especially Gaussian Processes (GP) (\citet{RasmussenWilliams2006}), to keep a memory of explored points with the aim to explore decision space while keeping a parsimonious evaluation budget. First introduced by \citet{Mockus1978} and further developed and popularized by \citet{Jones1998} in the noise free setting, it has latter been extended to stochastic black box optimisation (\citet{Frazier2009, Frazier2018}; \citet{Srinivas2009}; \citet{Picheny2013benchmark}; \citet{Hernandez2014};  \citet{Jalali2017} and references therein) and further sequential strategies have been studied (\citet{Risk2018}; \citet{Binois2019}). However, existing approaches typically assume Gaussian response distributions $\distf \X $. 
	Another branch of study pertaining to geostatistics and that does not rely on Gaussian or specific distributional assumptions is the so-called distributional Kriging 
	(\citet{Aitchison1982};  \citet{Egozcue2006};  \citet{Talska2018}) but such approaches are ill-suited in the case of moderate sample size heterogeneously scattered across space.
	
	\quad In this paper, we investigate and leverage a non-parametric Bayesian model for fields of probability distributions that generalizes the logistic Gaussian process model. We focus on the situation where, for each $\X$, the distribution $\distf \X $ admits a density $\densf \X$ with respect to a prescribed dominating measure. 
	Our approach is adapted to the case of moderate and heterogeneously scattered across space sample size. It delivers a probabilistic prediction of the field of probability distributions which allows us to derive sampling acquisition schemes. We introduce the aforementioned model in the Section \ref{sec:1}. Then, we study an application of this model in sequential design in Section \ref{sec:2} and demonstrate its applicability on toy probability density fields and on a contaminant localization problem in Section \ref{sec:3}.
	
	\quad The main contribution of this paper lies in the exploitation of a spatial extension of the logistic Gaussian process model for the sequential design of stochastic simulations towards optimisation and inversion. 
	
	\section{Sample-based modelling of density valued fields}
	\label{sec:1}
	
	\subsection{The logistic Gaussian process for density estimation}
	\label{subsec:1.1}
	\quad Before considering the full scale problem of density field estimation, let us briefly consider the simpler problem of (univariate) density estimation without any indexing on a spatial variable. 
	
	\quad Our focus is on a class of Bayesian density estimation methods where prior distributions of probability density functions are constructed by rescaling positive random processes in such a way that resulting realisations integrate to one. When the considered random processes are obtained by taking the logistic density transform (exponentiation and normalisation) of a Gaussian process, the resulting model is called Logistic Gaussian Process (LGP).
	LGP for density estimation were established and studied by \citet{Leonard1978} and \citet{Lenk1988, Lenk1991}. More recently, \citet{TokdarConsist2007} demonstrated that a large class of LGP achieves strong and weak consistency (under some regularity assumptions) at true densities that are either continuous or piecewise continuous.
	
\begin{figure}[!h]
		\centering
		\includegraphics[width=0.6\linewidth]{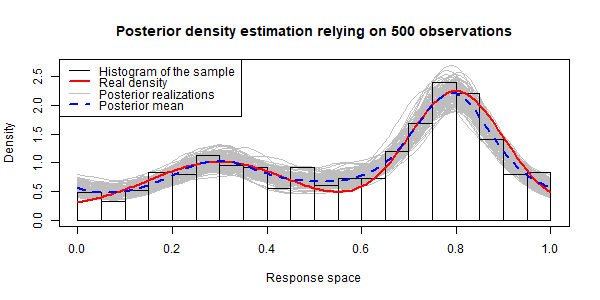}
		\caption{Using the logistic Gaussian process model in density estimation.}
		\label{fig:LGP_dens_est}
	\end{figure}

	\quad Using this prior within a Markov Chain Monte Carlo framework allows approximate sampling from the posterior distribution over the space of densities on $\I$, as illustrated in Figure~\ref{fig:LGP_dens_est}.
	
	\quad Models relying on other mappings than the \textit{logistic density transform} have been introduced and studied. In the particular case where the sigmoid (\textit{logistic function}) is used instead of an exponential, \citet{adams2009gaussian} introduced an exact sampling approach relying on the boundedness of the sigmoid. More recently, \citet{donner2018efficient} proposed an augmented model that allows for efficient inference by Gibbs sampling and an approximate variational mean field approach. However, our first experiments hinted that choosing a bounded function such as the sigmoid often leads to a lack of expressiveness, hence we decided to rely on LGP settings.
	
	\subsection{A spatial extension of the logistic Gaussian process for density field estimation}
	\label{subsec:1.2}
	\quad When modelling and optimising functions depending on both decision and uncontrolled variables, it can be a relevant option to use augmented models incorporating the uncontrolled variables as part of their inputs. In \citet{Picheny2013benchmark}, a GP indexed by the joint space of design parameters and computational time is considered to model partially converged simulation data. In \citet{Janusevskis2013}, a GP model over the joint spaces of controllable and uncontrollable variables is used for simulation-based robust optimisation.
	
	\quad Here we consider a spatial extension of the logistic Gaussian process. We work with a GP indexed by the product space $D \times \I$ and apply a logistic density transform to each $\X$-slice of the process to produce a probability density field.  A similar spatial LGP was introduced in \citet{tokdar2010} in the context of conditional density modelling (i.e., with random $\X$). To our knowledge, this family of models has not yet been used in the present context of sequential design of experiments, be it for stochastic optimisation or towards other goals. 
	
	\quad Here we assume that our probability measures $\distf \X $ of interest possess probability densities $\densf \X$ (with respect to some prescribed reference measure(s)) and we appeal to the said generalization of LGP to model the resulting density field. The resulting Spatial Logistic Gaussian process (SLGP) is defined by  
	\begin{equation}
		\label{eq:Spatial_LGP}
		\slgp_\X(t) =\dfrac{e^{W_{\X, t}}}{ \int_\mathcal I e^{ W_{\X, u} } \,du}
		\hspace{1cm} 
		(\X\in D, t \in \mathcal{I}),
	\end{equation}
	where $W_{\X, t} \sim \mathcal{GP}(\gpmean, k)$ is a measurable GP with given mean function $\gpmean:D \times \mathcal I \rightarrow \xR$ and covariance kernel $k: (D\times \mathcal I ) \times (D \times \mathcal I) \rightarrow \xR$ such that for any $\X \in D$, 
	$\int_\I e^{ W_{\X, u}} \,du < \infty$ almost surely.
	
	\begin{rmrk} When the index space $D$ is a singleton, SLGP coincides with the LGP mentioned in Section~\ref{subsec:1.1}.
	\end{rmrk}
	\begin{rmrk} Throughout the paper, the Lebesgue measure is assumed as dominating one, $\mathcal{I}$ is $[0,1]$, and the considered GPs possess smoothness properties amply guaranteeing the a.s. existence of the considered integrals.     
	\end{rmrk}
	
	\subsection{Conditioning on data and practicalities}
	\label{subsec:1.3}
	\quad In the SLGP model, the denominator involves values of $W$ over the whole domain. This infinite dimensional objects makes likelihood-based computations cumbersome. In \citet{TokdarTowards2007}, this problem is reduced to the one of a finite dimensional integral by relying on a moderate number of inducing points, introduced to reduce the dimensionality of the LGP. These inducing points are selected according to a life and death process. 
	
	\quad Here, we opt for another way to reduce the dimensionality that is more suitable for sequential optimisation. We consider finite-rank GP with the following parametrisation
	\begin{equation}
		\label{eq:GPextension}
		\ W_{\X, t} = \gpmean(\X,t) + \sum_{j=1}^p \sqrt{\lambda_j} e_j(\X, t) \varepsilon_j \ (\X \in D, \ t\in \mathcal I )
	\end{equation}
	where $p \in \xN$, the $e_i$'s are functions, the $\lambda_i$'s are scalars satisfying $\lambda_1 \geq \hdots \geq \lambda_p > 0$ and the $\varepsilon_i$'s are i.i.d. $\mathcal{N}(0, 1)$. Truncated Karhunen-Loève expansions constitute an important class falling under this framework. This simple form of the GP allows us to pre-calculate the values of $e_j$'s at the grid points and then swiftly obtain evaluations of $W$ and its integral (through a quadrature) for varying realisations (or tentative values) $\epsilon_i$'s of the $\varepsilon_i$'s. 
	
	\quad Now, consider that our data consist in  $n$ couples of locations and observations $\{(\X_i, t_i)\}_{1 \leq i \leq n}\in ( D\times \I )^n$. Moreover, assume the $t_i$'s are realisations of some independent random variables $T_i$, and let $\distf{\X_i} $ be the (unknown) distribution of $T_i$. We will note $\mathbf{T}_n = (T_i)_{1 \leq i \leq n}$ and $\mathbf{t}_n = (t_i)_{1 \leq i \leq n}$ .
	The GP $W$ defined by the Equation~\ref{eq:GPextension}, as well as the associated SLGP inherit their randomness from the Gaussian random vector $\pmb{\varepsilon}=(\varepsilon_i)_{1\leq i \leq p}$. 
	To highlight this dependency we shall also denote the SLGP $\slgp_\X(\cdot)$ by $\slgp_\X(\cdot ; \pmb{\varepsilon})$ and use the notation $\slgp_\X(\cdot ; \pmb{\epsilon})$ when instantiating the SLGP with a fixed value $\pmb{\epsilon}$ of $\pmb{\varepsilon}$. We perform a Bayesian estimation of the true unknown density field $\densf{\X_i}$ in the following way:
	
	\begin{itemize}
		\item  We use the prior $\pmb{\varepsilon}=(\varepsilon_i)_{1\leq i \leq p} \sim \mathcal{N}(\mathbf 0 , \mathbf{I}_{p})$, with density $\phi(\cdot)$ (with respect to Lebesgue measure in $\mathbb{R}^p$).	
		\item The finite-rank SLGP $\slgp_\X(\cdot ; \pmb{\varepsilon}):=\dfrac{e^{\gpmean(\X, \cdot) + \sum_{j=1}^p \sqrt{\lambda_j} e_j(\X, \cdot) \varepsilon_j}}{\displaystyle \int_\mathcal I e^{\gpmean(\X, u) + \sum_{j=1}^p \sqrt{\lambda_j} e_j(\X, u) \varepsilon_j} \,du} $ is used to model $\densf{\X}(\cdot)$.\\
		Hence, for $T_1, ..., T_n$ independent responses at locations $\mathbf x_1,..., \mathbf x_n$, one obtains the likelihood  \begin{equation}
			\label{eq:likeSLGP}
			L(\cdot; \mathbf{t}_n):
			\pmb{\epsilon}\in \xR \to  L(\pmb{\epsilon}; \mathbf{t}_n) = \prod\limits_{i=1}^n \slgp_{\X_i}(t_i ; \pmb{\epsilon})
		\end{equation}
		
		\item The posterior density of $ \pmb{\varepsilon}$, given by Bayes formula is $
		\pi( \pmb{\epsilon} \vert \mathbf T_n = \mathbf t_n ) \propto \phi( \pmb{\epsilon}) \prod_{i=1}^n \slgp_{\X_i}(t_i ; \pmb{\epsilon})$
	\end{itemize}
	
	\quad We leverage the fact that the posterior is known (up to a constant) to implement our density field estimation via a preconditioned Crank Nicholson algorithm (\citet{Cotter2013}). This approach delivers a probabilistic prediction of $(\densf \X )_{\X \in D}$, and allows us to approximately sample from the SLGP posterior distribution and recommend new sampling locations. 
	
	\section{Some first contribution in stochastic optimisation}
	\label{sec:2}
	
	\quad One of the main challenges in optimisation is to define suitable sequential design strategies. Choosing where to add observations is indeed crucial and a good design strategy must achieve a trade-off between exploration and exploitation of the input space so as to discover regions of interest while avoiding getting trapped in the vicinity of local minimisers or of artefactual basins of attraction. 
	
	\quad Deriving such strategies requires anticipating (probabilistically) the effect of adding new observations. To this end, meta-models are commonly used. We now present the considered optimisation problem, the criterion used to derive a sequential design strategy, and a stochastic simulation-based approach to estimate this criterion.
	
	\subsection{Optimisation problem considered}
	\label{subsec:2.2}
	\quad We recall that we are interested in minimising $ g(\X) = \rho(\distf\X )$, while we do not know the field $\distf\X $ (or equivalently its density $\densf\X$). We consider $G_\X$, the random function obtained by applying $\rho$ to the density valued field delivered by the SLGP model.  $G_\X$ is the model for $g(\X)$ induced by the SLGP $\slgp_\X(\cdot)$.
	
	\quad In the rest of the paper we will consider the particular case where $\rho$ is the median, but the presented approach is not restricted to this choice and can be applied to arbitrary (measurable) mappings, potentially also mappings depending on $\X$. 
	
	\begin{rmrk} 
		$G_\X$ remains uncertain knowing $\mathbf{T_n} = \mathbf{t_n}$ because of the conditional variability of $\pmb \varepsilon$. 
	\end{rmrk}
	
	\quad In the spirit of robust optimisation, we will consider the problem of minimising an $\alpha$-quantile of the random function $G_\X$. Here the value of $\alpha$ is arbitrarily set to 0.9 and stays fixed through the optimisation procedure. We note $Q_{n}(\X)$ the $\alpha$-quantile of $G_\X$. $Q_{n}(\X)$ is a random function as the observations $\mathbf T_n$ are left in random form. We shall also consider $Q_{n+K}(\X; \X_{\text{new}})$, the $\alpha$-quantile of $G_\X$ conditioned on past observations $T_n$ and on a batch of $K$ i.i.d. observations to be made at $\X_{\text{new}} \in D$.
	Denoting $\mathbb{E}_n[\cdot] = \mathbb{E}[\cdot \vert \mathbf{T_n}=\mathbf{t_n}]$, we consider:
	
	\begin{equation}
		\label{eq:EQI}
		\text{EQI}_{n}(\X_{\text{new}},K) 
		= \mathbb{E}_n 
		\left[
		\left(
		\min_{\X \in D} Q_{n}(\X) 
		- 
		\min_{\X \in D}  Q_{n+K}(\X; \X_{\text{new}})  \right)^+\right].
	\end{equation}
	
	\begin{rmrk} This criterion was inspired by the Expected Quantile Improvement presented in \citet{Picheny2013quantile}, which would write here as $\displaystyle \mathbb{E}_n [( \min_{i\leq n} Q_{n}(\X_i)- Q_{n+1}(\X_{\text{new}};\X_{\text{new}}) )^+]$, but is modified in the spirit of knowledge gradient approaches from \citet{Frazier2009} to account for improvements on the whole domain.
	\end{rmrk}
	
	\subsection{Simulation-based computation of criteria}
	\label{subsec:2.3}
	
	\quad Classically, in Sequential Uncertainty Reduction (SUR) (\citet{Bect2019}) approaches, it is assumed that the function of interest is a realisation of a GP. Under these assumptions, several criteria enjoy (semi-)analytical forms, favouring criterion optimisation and the implementation of design strategies. \\
	However, in our situation, it does not appear feasible to obtain a closed-form formula for the considered EQI criterion and we therefore estimate it via stochastic simulation.  
	
	\quad In order to quantify the effect of adding an observation $T_{n+1}$ at a given location $\X_{\text{new}}$, one needs to study the probability density of a new observation $T$ at $\X$ conditioned on $\mathbf{T_n}=\mathbf{t_n}$. 
	In favour of the law of total probability, and by considering Equation \ref{eq:likeSLGP}, one finds that it is given by:
	\begin{equation}
		\label{eq:Sim_Current_Density}
		\pi(t \vert \mathbf{T_n}=\mathbf{t_n}) \propto \int \pi(t \vert \pmb \epsilon) \pi(\pmb \epsilon \vert \mathbf{t_n}) \,d\pmb \epsilon
		\propto \int \slgp_\X(t ; \pmb{\epsilon}) \pi(\pmb \epsilon \vert \mathbf{t_n}) \,d\pmb\epsilon .
	\end{equation}
	\quad This motivates the following approach, which can be considered as a basic application of Sequential Monte Carlo (\citet{Doucet2013}), where we use a  simple simulation-based particle filter to approximate an unknown future quantity:
	
	\begin{enumerate}
		\item The generative model given by the SLGP model is implemented within a MCMC framework and used to obtain $N$ approximate realisations of $ \pmb \varepsilon \vert \mathbf T_n = \mathbf t_n$, noted $( \pmb \epsilon_{(j)} )_{1 \leq j \leq N }$. \\
		The density of a new observation at $\X$  (See Equation~\ref{eq:Sim_Current_Density}) is approximated by the mixture $ N^{-1}  \sum\limits_{j=1}^N  \slgp_{\X}(\cdot ; \pmb\epsilon_{(j)} )$.
		
		\item The impact of adding $K$ observations at a given location $\X_{\text{new}}$ is estimated by doing $M$ simulations:\\
		For the $i$-th simulation, $K$ realisations of the random variable $\widetilde{T}_{\text{new}}$ are independently drawn from the density $N^{-1} \sum\limits_{j=1}^N  \slgp_{\X_{\text{new}}}(\cdot ; \pmb\epsilon_{(j)} )  $. The corresponding  batch of response values is noted $\widetilde{\mathbf{t}}^{(i)}_{\text{new}} = (\widetilde{t}^{(i), 1}_{\text{new}}, ..., \widetilde{t}^{(i), K}_{\text{new}})$.\\
		In light of Equation~\ref{eq:Sim_Current_Density}, the response density at $\X$  conditional on past data and on the simulated batch is given by 
		$\pi(t \vert \mathbf{T_n}=\mathbf{t_n}, \widetilde{\mathbf{T}}^{(i)}_{\text{new}}=\widetilde{\mathbf{t}}^{(i)}_{\text{new}}  ) \propto \displaystyle \int \slgp_\X(t ; \pmb{\epsilon})  \prod\limits_{\ell=1}^K \slgp_{\X_{\text{new}}}( \widetilde{t}^{(i), \ell}_{\text{new}} ; \pmb{\epsilon}) \pi(\pmb \epsilon \vert \mathbf{t_n}) \,d\pmb \epsilon $, leveraging the fact that the $\pmb \epsilon_j$'s are drawn from the posterior, we can approximate the integral by the Monte Carlo sum $\sum\limits_{j=1}^N  \slgp_\X( \cdot ; \pmb{\epsilon}_{(j)}) \omega_{i, j}$, with weights $\omega_{i, j}$ proportional to $\prod\limits_{\ell=1}^K \slgp_{\X_{\text{new}}}( \widetilde{t}^{(i), \ell}_{\text{new}} ; \pmb{\epsilon}_{(j)})$ and summing to one.
		\item 
		The density field distribution after adding $K$ observations at $\X_{\text{new}}$ is predicted by 
		$M^{-1} \sum\limits_{i=1}^M \sum\limits_{j=1}^K  \slgp_{\X}(\cdot ; \pmb{\epsilon}_{(j)}) \omega_{i, j}$.
	\end{enumerate}
	
	\section{Applications}
	\label{sec:3}
	
	\quad For all the coming applications, we consider a zero mean GP $W_{\X, t} = \sum_{j=1}^p \sqrt{\lambda_j} e_j(\X, t) \varepsilon_j, \ p \in \xN$, with $t$ and $\X$ being uni-variate. To ensure consistency with the rest of the document, we will keep the bold notation for $\X$. To define the $e_j$'s, we start from bi-variate Fourier functions of order $q > 0$: sine and cosine of $2\pi( \omega_1 t + \omega_2 \X)$, where $ \omega_1$ and $\omega_2$ are integers satisfying $-q \leq  \omega_1, \omega_2 \leq q$. Then, we remove redundant terms as well as those irrelevant in the SLGP setting (i.e. functions independent of $t$, that would be simplified with the normalisation of the process) and set the corresponding $\lambda_j^2$ to be $1 / (1+ \omega_1 + \omega_2 )$. 
	
	\subsection{Some analytical applications}
	\label{subsec:3.1}
	
\begin{figure}[H]
		\centering
		\includegraphics[width=0.8\linewidth]{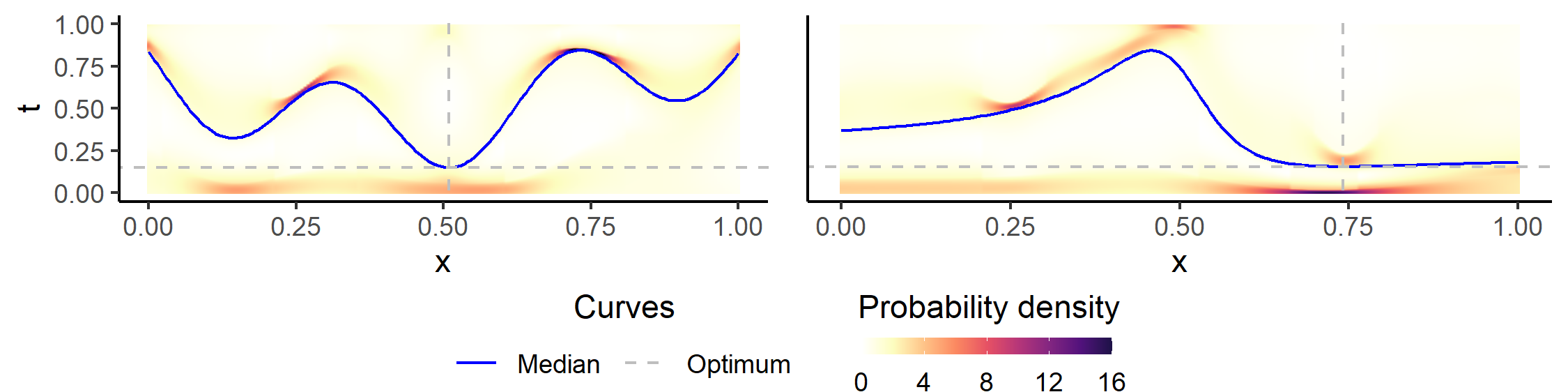}
		\caption{The two multi-modal reference density fields, their median and its global optimum.}
		\label{fig:Ref_fun_N2}
	\end{figure}

	\quad In the analytical applications, we have $D=\I=[0, 1]$ and consider four known density fields. The median functions of these fields appear in Figure \ref{fig:Ref_fun_N2} and are defined as $f_1(\X) = 0.25\sin(16\X+9) + 0.25 \sin(4.8 \X + 2.7) + 0.625$ (minimum at $\X^* \approx 0.5095$) and $f_2(\X) = 0.15 + \frac{7}{72} \frac{1.1(10\X-5)^2-5(10\X-5) + 6.1}{(10\X-5)^2 +1 }$ (minimum at $\X^* \approx 0.7414$).
	
	\quad Our first class of probability density fields, that we will refer to as ``truncated Gaussian fields'', writes as $h_\X \left(\frac{t-f_i(\X)}{0.05}\right)$, $i\in\{1, 2\}$, with $h_\X$ a symmetrically truncated standard Gaussian with thresholds $\pm \min\left(f_i(\X),1-f_i(\X)\right)$. The truncation ensures that the distribution remains symmetrical around its median and mean $f_i$. The second case - that we will refer to as ``multi-modal field''- is of the same form yet with $h_\X$ median-0 but multi-modal such as represented in Figure~\ref{fig:Ref_fun_N2}.

\begin{figure}[H]
		\centering
		\includegraphics[width=0.8\linewidth]{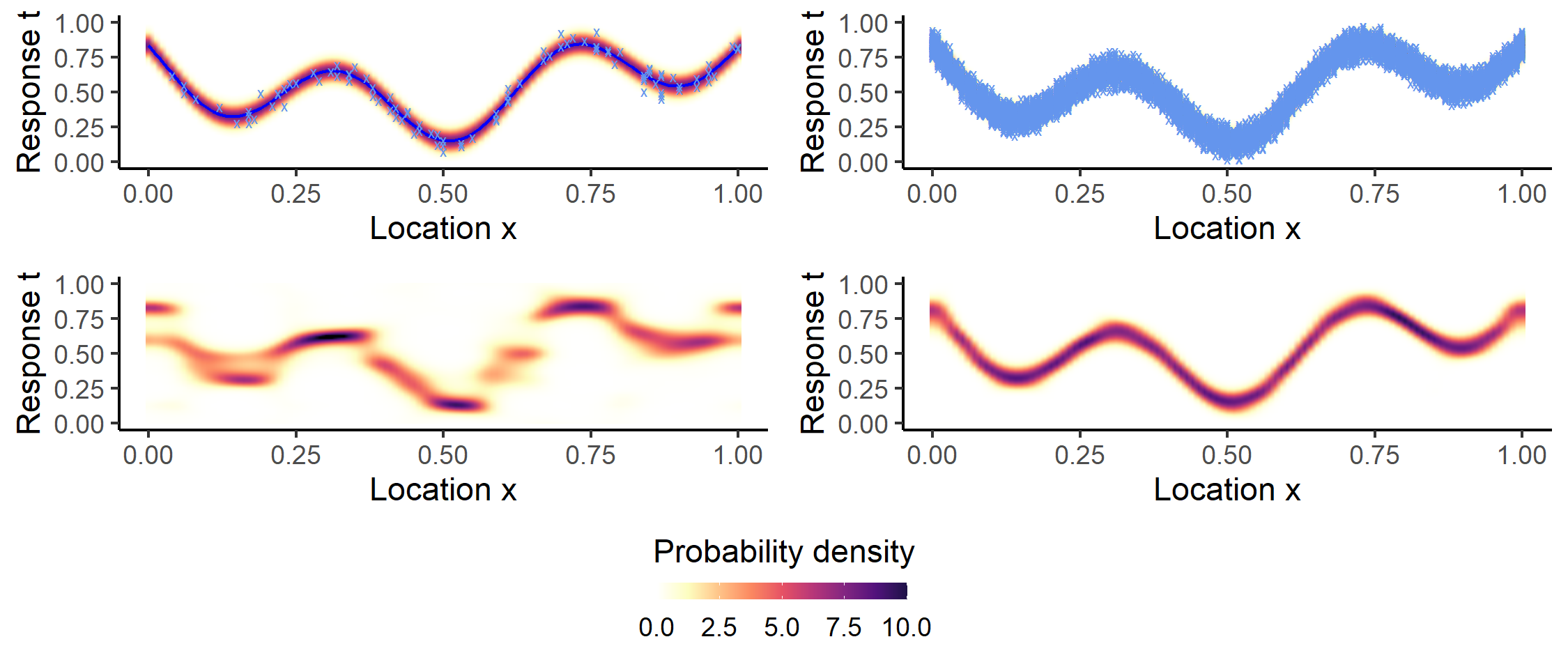}
		\caption{Results the truncated Gaussian field with median $f_1$, using 121 basis functions. True field and samples used (top), posterior mean field (bottom) for a respective sample size of 100 (left) and 10000 (right).}
		\label{fig:Goodnessoffit}
	\end{figure}

	\quad We perform density field estimation as presented in Section \ref{subsec:1.3} for such reference fields for different sample sizes and order of the GP. In this section, we are not yet in an adaptive setting: each time a new observation is added to the model, its location is determined randomly, with uniform distribution over the index set. Figure~\ref{fig:Goodnessoffit} displays the posterior mean field with two sample sizes for the truncated Gaussian field with median $f_1$.
	\quad We observe in Figure~\ref{fig:Goodnessoffit} that a higher sample size seems to yield a better estimation. In order to quantify the prediction error for different sample sizes and GP's order, we define an integrated squared Hellinger distance to measure dissimilarity between two probability density valued fields $p^i=(p_{\X}^{i})_{\X \in D}$ ($i=1,2$):
	\begin{equation}
		d_{\text{ISH}}^2(p^{1} , p^{2}) = \int_D \int_\I \left( \sqrt{p_{\mathbf{v}}^{1}(u)} - \sqrt{p_{\mathbf{v}}^{2}(u)}  \right)^2 \,du \,d\mathbf{v}
	\end{equation} 
	
\begin{figure}[H]
		\centering
		\includegraphics[width=0.8\linewidth]{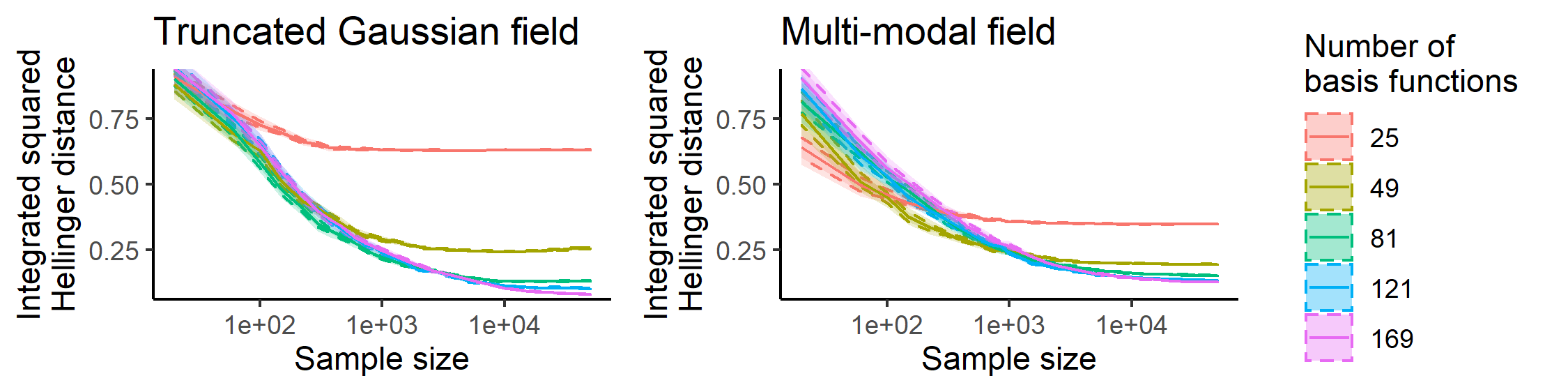}
		\caption{Integrated squared Hellinger distance distribution for different sample sizes and process orders, when the reference field has $f_1$ as its median.}
		\label{fig:IntegratedDH}
	\end{figure}

	\quad In Figure~\ref{fig:IntegratedDH}, we display the distribution of $d_{\text{ISH}}^2$ between true and estimated fields for various sample sizes and SLGP orders. As both functions yielded close results, we show only the results for $f_1$. We see that the errors are comparable for small sample sizes. The order becomes limiting when more observations are available as those of the considered SLGPs relying on the smallest numbers of basis functions appear to struggle capturing small scale variations.  
	
	\subsection{Test case: contaminant source localization under uncertain geology}
	\label{subsec:3.2}
\begin{figure}[h]
		\centering
		\includegraphics[width=0.95\linewidth]{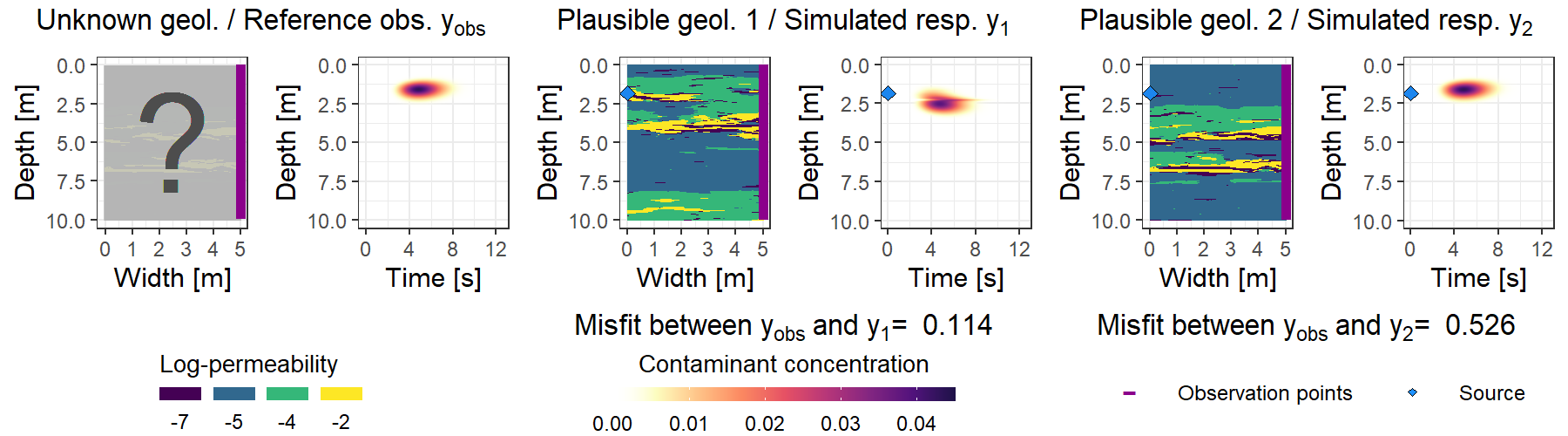}
		\caption{
		Geological medium versus breakthrough map in different situations. From left to right: 
		the first pair of plots depicts the starting situation our inverse problem of recovering the source location given an observed breakthrough map (right) yet under unknown geology (left). The second and third pairs of plots represent simulated breakthrough maps under randomly generated geologies, and the associated misfit values with respect to the observed breakthrough map. For simplicity, the same source depth (1.84m) is assumed in all three cases.
		}
		\label{fig:illu_geol}
	\end{figure}

\begin{figure}[h]
		\centering
		\includegraphics[scale=0.8]{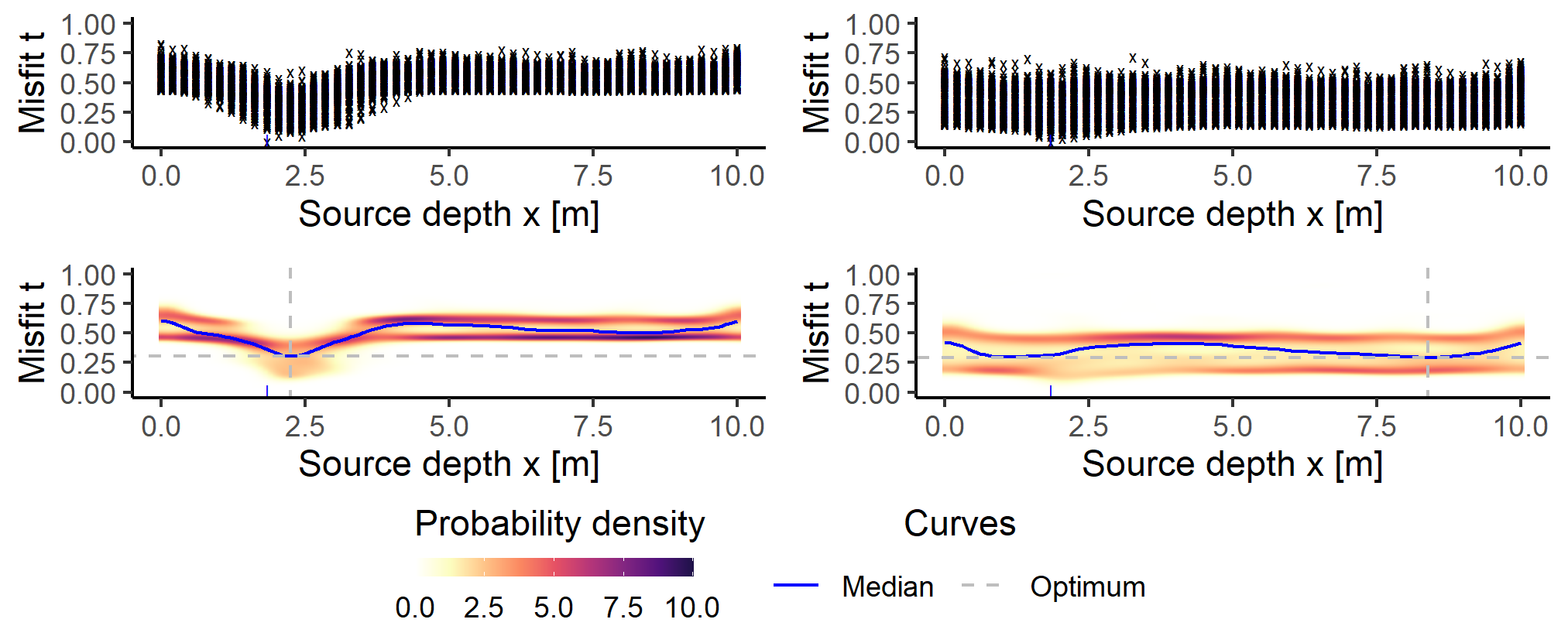}
		\caption{Misfit data (top) and posterior mean field (bottom) for two latent geological structures. }
		\label{fig:ref_geol}
	\end{figure}

	\quad 
	We now consider a hydro-geological test case in which, given breakthrough curves at several monitoring depths, we want to localize the source depth of a contaminant injected in a saturated aquifer whose geological properties are highly uncertain. We use the term \textit{breakthrough map} to speak of the stacked breakthrough curves at prescribed monitoring depths. Also, denote by \textit{observed} the given breakthrough map, even if throughout the section these observed maps stem from simulations with prescribed source depth and geology.
	\quad This inference is performed by measuring the dissimilarity (\textit{misfit}) between observed and simulated breakthrough maps, and our goal here is to uncover the depth that yields simulations with minimal median misfit. Here the index $\X$ stands for a candidate source depth while the response $t$ represents the dissimilarity between simulated and observed breakthrough maps. The stochasticity of the response is due to our imperfect knowledge of the geological properties of the aquifer, as sketched on Figure~\ref{fig:illu_geol} and further detailed below. Figure~\ref{fig:ref_geol} shows the misfit values obtained by comparing 10000 simulations (200 replications at 50 different source depth values) to two different observed breakthrough maps, as well as the posterior mean field obtained with a SLGP of order 121.  
	
	\quad To perform a flow simulation with a source at depth $\X$ while accounting for the uncertainty regarding the geological properties of the domain, we rely on plausible geological realisations. To keep the problem simple, the zone of interest of the aquifer is modelled as a 2D vertical section (10 meter deep and 5 meter wide) aligned with the main flow direction. For each simulation performed, the plausible geological realisation used is a multiple-point statistics realisation generated with the Deesse algorithm (\citet{Mariethoz2010}) that reproduce the complex features of braided-river aquifer models (\citet{Pirot2015}). Then, the contaminant flow and transport is simulated under steady-state flow and fixed boundary conditions (constant hydraulic gradient) using the Maflot Matlab code (\citet{Kunze2011}).
	
	\quad After running such a simulation for a contaminant released at time $0$, width $0$ and depth $\X$, we obtain a concentration breakthrough map (such as depicted in Figure~\ref{fig:illu_geol} under three different scenarios). Here $100$ candidate observation depths, regularly spaced between 0m and 10m, are considered. Each breakthrough curve, i.e., section of a breakthrough map at a a prescribed depth, describes the contaminant concentration observed at width 5m for varying times (discretised over 101 points regularly spaced between 0s and 12.5s). Misfits between observed and simulated curves are defined here in terms of a re-scaled 2-norm distance. 
	
	\quad 
	Throughout the section and as illustrated in Figure~\ref{fig:ref_geol}, we use two reference fields of misfit distributions that are obtained by fitting SLGP models to misfit data produced under the two latent geologies of Figure~\ref{fig:illu_geol}.  
	
\begin{figure}[h]
		\centering
		\includegraphics[scale=0.8]{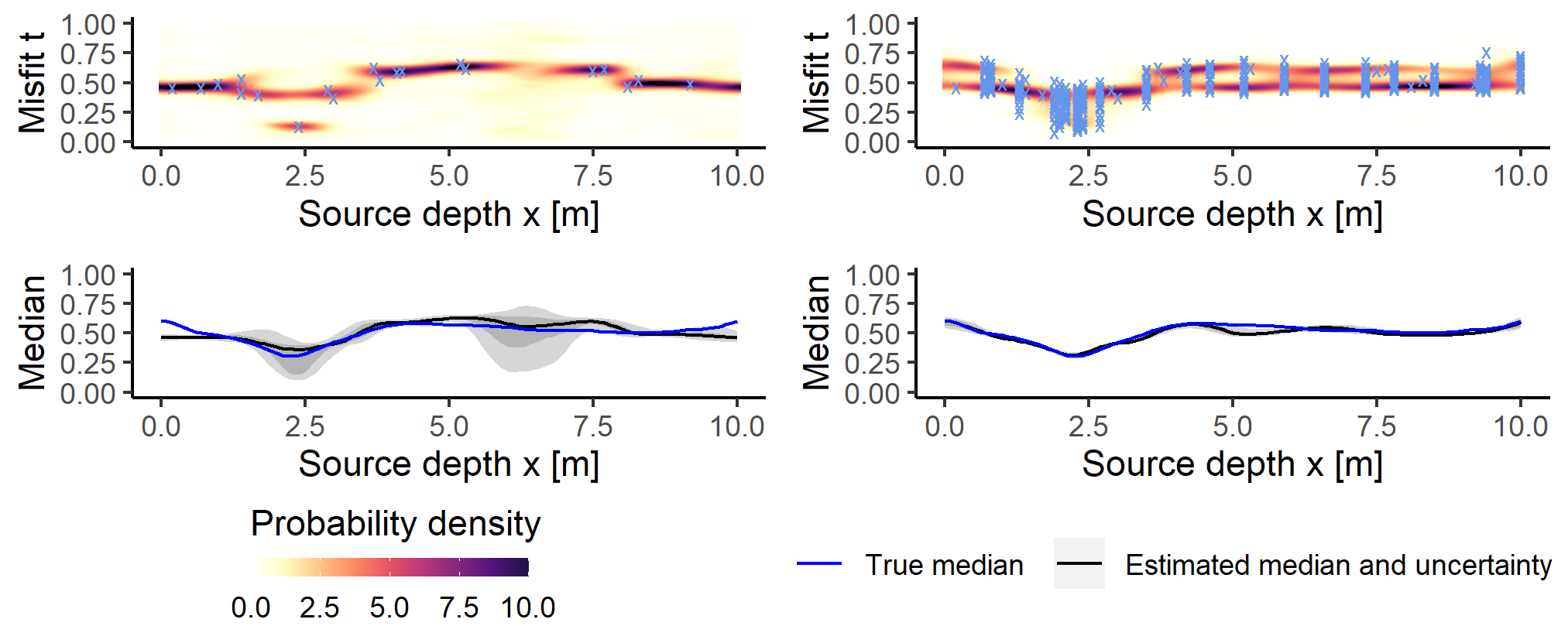}
		\caption{Results at the beginning of the algorithm [left] and after the 25th step [right].\\
			Mean field estimated and samples available [top]; Estimated VS reference median [bottom].}
		\label{fig:geol1}
	\end{figure}

	\quad Using these two reference density fields to draw new samples, we follow the methodology introduced in part \ref{sec:2} and represent in Figure \ref{fig:geol1} the posterior mean field and its estimated median before and after running 25 steps of the algorithm on the first geological structure. 
	
	\quad In these settings, the global minimum (at 2,24m) is easily found and the algorithm focuses on improving its estimation of the median. This corresponds to an exploitation-oriented approach. Although not displayed here, we found out, as reflected in Figure~\ref{fig:benchmark} that with the other geological structure, our approach was also able to locate the minimum, this time by focusing on exploring by adding observations at new locations of interest.
	
	\subsection{Optimisation benchmark}
	\label{subsec:3.3}
	
	\quad For our small benchmark, the starting design consists in $n = 20$ data points $(\X_i, t_i)$ from the reference fields heterogeneously scattered across space. Their location are independent uniformly distributed across parameter space. At each step, observations are added in batches of $20$ at the same location. We repeat 24 independent instances of the optimisation process for each strategy and each application and compare the performances in term of optimality gap (difference between real and estimated optimal medians). This approach is favoured due to the relatively high cost of one evaluation of the EQI criterion for SLGP, but we expect it to be detrimental to our GP-based competitors, as GPs would benefit more from having scattered observations rather than batches scattered over different points.\\
	
\begin{figure}[H]
		\centering
		\includegraphics[width=0.88\linewidth]{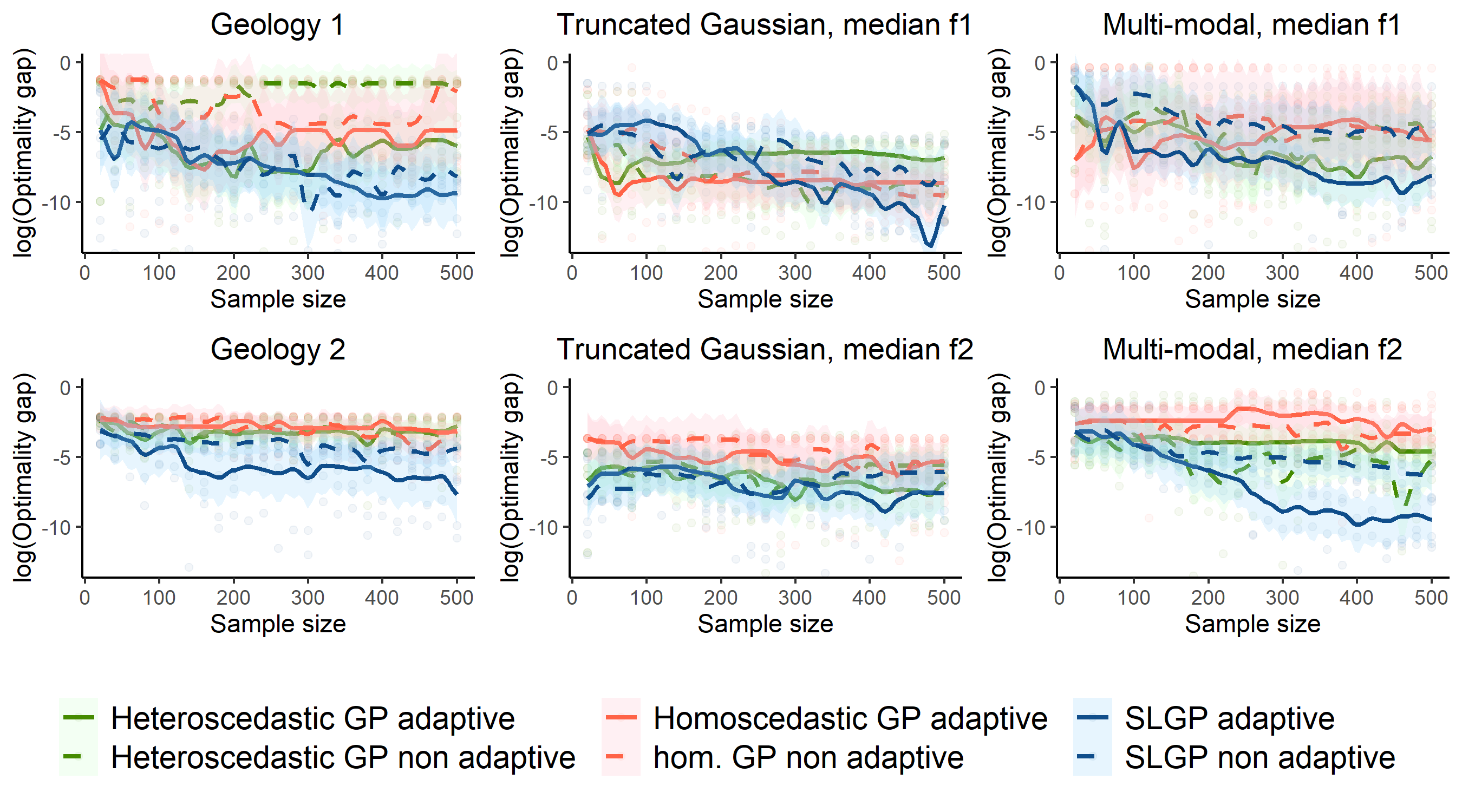}
		\caption{Median of the log optimality gap for the $6\times 6$ considered strategies and test cases.}
		\label{fig:benchmark}
	\end{figure}

	\quad  We compare different strategies for modelling the field and choosing the next sampling location.
	The value of the minimiser is inferred by modelling the function of interest with one of three models: the first one, that will be called homoscedastic GP consists in a GP regression where the observation noise level is assumed to be uniform throughout the domain. The second one, a GP regression with input dependent noise rates  as in \citet{kersting2007hetGP}, \citet{binois2018practical} will be called heteroscedastic GP. The last one is the SLGP model.\\ 
	For each of these three models, we compare a non-adaptive approach (at each step, new observations are added at a location chosen uniformly at random) to an adaptive approach. The criteria used are: Approximated Knowledge Gradient (AKG) as implemented in the R package DiceOptim version 2.0.1 \citep{DiceOptim} for the homoscedastic GP, the Expected improvement, as implemented in the R package hetGP version 1.2.1 \citep{hetGP} for the heteroscedastic GP, and the criterion \ref{eq:EQI} of part \ref{sec:2} for the SLGP.  The results of the benchmark are shown in Figure \ref{fig:benchmark}.
	
	\quad Hyper-parameters of the two GP's are estimated by maximum likelihood. For the adaptive SLGP, we decided to display the results obtained when minimising a 90\% quantile of the random median, as our first experiments showed no strong sensitivity to the chosen level $\alpha$. On a personal laptop with 16Go RAM and a simple R implementation, performing the SLGP density estimation took between 13 and 24 minutes depending on the sample size, and inferring the EQI with 150 simulations for 101 locations took 5 more minutes.
	
	\quad One notices that in the most complex situations, the sampling scheme based on GP modelling of the functional performs worse than the approaches based on SLGP modelling. We found out that GP based approaches tends to get trapped in local optima. This might be explained by the credit allocation.
	
	\section{Conclusion and perspectives}
	\label{conclusion}
	
	\quad We demonstrated how a spatial generalization of the Logistic Gaussian Process model can be used for sample-based modelling of distribution fields, and how the resulting probabilistic predictions of distribution fields can be leveraged for  moderate-dimensional stochastic optimization problems under unknown heterogeneous noise distributions. 
	In particular, we introduced a simple sequential Monte Carlo-based approach to quantify the impact of sampling at a location and showed, on a modest benchmark, that it enables implementing sequential design strategies trading off between exploration and exploitation.
	
	\quad The conducted experiments yielded promising results. However, further work is needed to more extensively compare algorithms with respect to the employed meta-models and criteria, respectively, and identify which of these degrees of freedom are most influential on performances. Tuning model settings and strategies appropriately could ultimately lead to substantial efficiency gains, be it towards stochastic optimization or broader uncertainty reduction goals. In particular, SLGP would benefit from more systematic model selection and parameter estimation approaches.
	
	\quad 
	Upcoming research directions encompass making the SLGP model more suitable for higher dimensions and reducing associated computational costs, assessing the performance of the resulting probabilistic predictions, and deriving further uncertainty reduction strategies. It is also of interest to investigate theoretical properties of SLGP models and considered sequential strategies, notably in view of underlying mean and covariance functions.
	
	\section*{Acknowledgements}
	The authors wish to thank the reviewing team for their remarks that helped substantially improve the paper. AG's and DG's contributions have taken place within the Swiss National Science Foundation project number 178858. AG and DG would like to warmly thank Dr. Tomasz Kacprzak (ETH Z\"urich) for early discussions having motivated part of this work, as well as Yves Deville for insightful comments. 
	
	\bibliographystyle{abbrvnat}

{\footnotesize
	\begin{thebibliography}{50}
		\providecommand{\natexlab}[1]{#1}
		\providecommand{\url}[1]{\texttt{#1}}
		\expandafter\ifx\csname urlstyle\endcsname\relax
		\providecommand{\doi}[1]{doi: #1}\else
		\providecommand{\doi}{doi: \begingroup \urlstyle{rm}\Url}\fi
		
		\bibitem[Aitchison(1982)]{Aitchison1982}
		J.~Aitchison.
		\newblock The statistical analysis of compositional data.
		\newblock \emph{Journal of the Royal Statistical Society, Series B:
			Methodological}, 44\penalty0 (2):\penalty0 139--177, 1982.
		
		\bibitem[Bect et~al.(2019)Bect, Bachoc, and Ginsbourger]{Bect2019}
		J.~Bect, F.~Bachoc, and D.~Ginsbourger.
		\newblock A supermartingale approach to {G}aussian process based sequential
		design of experiments.
		\newblock \emph{Bernoulli}, 25\penalty0 (4A):\penalty0 2883--2919, 2019.
		
		\bibitem[Bellini and Di~Bernardino(2017)]{Bellini2015}
		F.~Bellini and E.~Di~Bernardino.
		\newblock {R}isk management with expectiles.
		\newblock \emph{The European Journal of Finance}, 23\penalty0 (6):\penalty0
		487--506, 2017.
		
		\bibitem[Binois and Gramacy(2019)]{hetGP}
		M.~Binois and R.~B. Gramacy.
		\newblock \emph{hetGP: Heteroskedastic Gaussian Process Modeling and Design
			under Replication}, 2019.
		\newblock URL \url{https://CRAN.R-project.org/package=hetGP}.
		\newblock R package version 1.1.2.
		
		\bibitem[Binois et~al.(2018)Binois, Gramacy, and
		Ludkovski]{binois2018practical}
		M.~Binois, R.~B. Gramacy, and M.~Ludkovski.
		\newblock Practical heteroscedastic gaussian process modeling for large
		simulation experiments.
		\newblock \emph{Journal of Computational and Graphical Statistics}, 27\penalty0
		(4):\penalty0 808--821, 2018.
		
		\bibitem[Binois et~al.(2019)Binois, Huang, Gramacy, and Ludkovski]{Binois2019}
		M.~Binois, J.~Huang, R.~B. Gramacy, and M.~Ludkovski.
		\newblock Replication or exploration? {S}equential design for stochastic
		simulation experiments.
		\newblock \emph{Technometrics}, 61\penalty0 (1):\penalty0 7--23, 2019.
		
		\bibitem[Birge and Louveaux(2011)]{Birge2011}
		J.~R. Birge and F.~Louveaux.
		\newblock \emph{Introduction to stochastic programming}.
		\newblock Springer Science \& Business Media, 2011.
		
		\bibitem[Bubeck and Cesa-Bianchi(2012)]{Bubeck2012}
		S.~Bubeck and N.~Cesa-Bianchi.
		\newblock Regret analysis of stochastic and nonstochastic multi-armed bandit
		problems.
		\newblock \emph{Foundations and Trends{\textregistered} in Machine Learning},
		5\penalty0 (1):\penalty0 1--122, 2012.
		
		\bibitem[Cesa-Bianchi and Lugosi(2006)]{Cesa2006}
		N.~Cesa-Bianchi and G.~Lugosi.
		\newblock \emph{Prediction, learning, and games}.
		\newblock Cambridge university press, 2006.
		
		\bibitem[Cotter et~al.(2013)Cotter, Roberts, Stuart, and White]{Cotter2013}
		S.~Cotter, G.~Roberts, A.~Stuart, and D.~White.
		\newblock {MCMC} methods for functions: modifying old algorithms to make them
		faster.
		\newblock \emph{Statistical Science}, 28\penalty0 (3):\penalty0 424--446, 2013.
		
		\bibitem[Donner and Opper(2018)]{donner2018efficient}
		C.~Donner and M.~Opper.
		\newblock Efficient {B}ayesian inference for a {G}aussian process density
		model.
		\newblock In A.~Globerson and R.~Silva, editors, \emph{Proceedings of the
			Thirty-Fourth Conference on Uncertainty in Artificial Intelligence}, pages
		53--62. {AUAI} Press, 2018.
		
		\bibitem[Doucet et~al.(2001)Doucet, De~Freitas, and Gordon]{Doucet2013}
		A.~Doucet, N.~De~Freitas, and N.~Gordon.
		\newblock \emph{Sequential {M}onte {C}arlo methods in practice}.
		\newblock Springer, New York, NY, 2001.
		
		\bibitem[Efromovich(2010)]{Efromovich2010}
		S.~Efromovich.
		\newblock Dimension reduction and adaptation in conditional density estimation.
		\newblock \emph{Journal of the American Statistical Association}, 105\penalty0
		(490):\penalty0 761--774, 2010.
		
		\bibitem[Egozcue et~al.(2006)Egozcue, Díaz–Barrero, and
		Pawlowsky-Glahn]{Egozcue2006}
		J.~J. Egozcue, J.~Díaz–Barrero, and V.~Pawlowsky-Glahn.
		\newblock {H}ilbert space of probability density functions based on {A}itchison
		geometry.
		\newblock \emph{Acta Mathematica Sinica, English Series}, 22\penalty0
		(4):\penalty0 1175--1182, 2006.
		
		\bibitem[Frazier et~al.(2009)Frazier, Powell, and Dayanik]{Frazier2009}
		P.~Frazier, W.~Powell, and S.~Dayanik.
		\newblock The knowledge-gradient policy for correlated normal beliefs.
		\newblock \emph{INFORMS journal on Computing}, 21\penalty0 (4):\penalty0
		599--613, 2009.
		
		\bibitem[Frazier(2018)]{Frazier2018}
		P.~I. Frazier.
		\newblock A tutorial on {B}ayesian optimization.
		\newblock \emph{arXiv preprint arXiv:1807.02811}, 2018.
		
		\bibitem[Hall et~al.(2004)Hall, Racine, and Li]{Hall2004}
		P.~Hall, J.~Racine, and Q.~Li.
		\newblock Cross-validation and the estimation of conditional probability
		densities.
		\newblock \emph{Journal of the American Statistical Association}, 99\penalty0
		(468):\penalty0 1015--1026, 2004.
		
		\bibitem[Hern{\'a}ndez-Lobato et~al.(2014)Hern{\'a}ndez-Lobato, Hoffman, and
		Ghahramani]{Hernandez2014}
		J.~M. Hern{\'a}ndez-Lobato, M.~W. Hoffman, and Z.~Ghahramani.
		\newblock Predictive entropy search for efficient global optimization of
		black-box functions.
		\newblock 27:\penalty0 918--926, 2014.
		
		\bibitem[Jalali et~al.(2017)Jalali, Nieuwenhuyse, and Picheny]{Jalali2017}
		H.~Jalali, I.~Nieuwenhuyse, and V.~Picheny.
		\newblock Comparison of {K}riging-based algorithms for simulation optimization
		with heterogeneous noise.
		\newblock \emph{European Journal of Operational Research}, 261\penalty0
		(1):\penalty0 279--301, 2017.
		
		\bibitem[Janusevskis and Le~Riche(2013)]{Janusevskis2013}
		J.~Janusevskis and R.~Le~Riche.
		\newblock Simultaneous kriging-based estimation and optimization of mean
		response.
		\newblock \emph{Journal of Global Optimization}, 55\penalty0 (2):\penalty0
		313--336, 2013.
		
		\bibitem[Jones et~al.(1998)Jones, Schonlau, and Welch]{Jones1998}
		D.~Jones, M.~Schonlau, and W.~Welch.
		\newblock Efficient {G}lobal {O}ptimization of expensive black-box functions.
		\newblock \emph{Journal of Global optimization}, 13\penalty0 (4):\penalty0
		455--492, 1998.
		
		\bibitem[Kersting et~al.(2007)Kersting, Plagemann, Pfaff, and
		Burgard]{kersting2007hetGP}
		K.~Kersting, C.~Plagemann, P.~Pfaff, and W.~Burgard.
		\newblock Most likely heteroscedastic gaussian process regression.
		\newblock In \emph{Proceedings of the 24th International Conference on Machine
			Learning}, ICML'07, pages 393--400, 2007.
		
		\bibitem[K\"{u}nze and Lunati(2011)]{Kunze2011}
		R.~K\"{u}nze and I.~Lunati.
		\newblock A matlab toolbox to simulate flow through porous media.
		\newblock Technical report, University of Lausanne, Switzerland, 2011.
		
		\bibitem[Lenk(1988)]{Lenk1988}
		P.~J. Lenk.
		\newblock The logistic normal distribution for {B}ayesian, nonparametric,
		predictive densities.
		\newblock \emph{Journal of the American Statistical Association}, 83\penalty0
		(402):\penalty0 509--516, 1988.
		
		\bibitem[Lenk(1991)]{Lenk1991}
		P.~J. Lenk.
		\newblock Towards a practicable {B}ayesian nonparametric density estimator.
		\newblock \emph{Biometrika}, 78\penalty0 (3):\penalty0 531--543, 1991.
		
		\bibitem[Leonard(1978)]{Leonard1978}
		T.~Leonard.
		\newblock Density estimation, stochastic processes and prior information.
		\newblock \emph{Journal of the Royal Statistical Society, Series B:
			Methodological}, 40\penalty0 (2):\penalty0 113--132, 1978.
		
		\bibitem[Mandt et~al.(2017)Mandt, Hoffman, and Blei]{Mandt2017}
		S.~Mandt, M.~D. Hoffman, and D.~M. Blei.
		\newblock Stochastic gradient descent as approximate bayesian inference.
		\newblock \emph{The Journal of Machine Learning Research}, 18\penalty0
		(1):\penalty0 1--35, 2017.
		
		\bibitem[Mariethoz et~al.(2010)Mariethoz, Renard, and
		Straubhaar]{Mariethoz2010}
		G.~Mariethoz, P.~Renard, and J.~Straubhaar.
		\newblock The direct sampling method to perform multiple-point geostatistical
		simulations.
		\newblock \emph{Water Resources Research}, 46\penalty0 (11):\penalty0 W11536,
		2010.
		
		\bibitem[Mo{\v{c}}kus et~al.(1978)Mo{\v{c}}kus, Tiesis, and
		{\v{Z}}ilinskas]{Mockus1978}
		J.~Mo{\v{c}}kus, V.~Tiesis, and A.~{\v{Z}}ilinskas.
		\newblock The application of {B}ayesian methods for seeking the extremum.
		\newblock \emph{Towards global optimization}, 2:\penalty0 117--129, 1978.
		
		\bibitem[Moutoussamy et~al.(2015)Moutoussamy, Nanty, and
		Pauwels]{Moutoussamy2015}
		V.~Moutoussamy, S.~Nanty, and B.~Pauwels.
		\newblock Emulators for stochastic simulation codes.
		\newblock \emph{ESAIM: Proceedings and Surveys}, 48:\penalty0 116--155, 2015.
		
		\bibitem[Murray et~al.(2009)Murray, MacKay, and Adams]{adams2009gaussian}
		I.~Murray, D.~MacKay, and R.~P. Adams.
		\newblock The gaussian process density sampler.
		\newblock In \emph{Advances in {N}eural {I}nformation {P}rocessing {S}ystems},
		volume~21, pages 9--16, 2009.
		
		\bibitem[Picheny et~al.(2013{\natexlab{a}})Picheny, Ginsbourger, Richet, and
		Caplin]{Picheny2013quantile}
		V.~Picheny, D.~Ginsbourger, Y.~Richet, and G.~Caplin.
		\newblock Quantile-based optimization of noisy computer experiments with
		tunable precision.
		\newblock \emph{Technometrics}, 55\penalty0 (1):\penalty0 2--13,
		2013{\natexlab{a}}.
		
		\bibitem[Picheny et~al.(2013{\natexlab{b}})Picheny, Wagner, and
		Ginsbourger]{Picheny2013benchmark}
		V.~Picheny, T.~Wagner, and D.~Ginsbourger.
		\newblock A benchmark of kriging-based infill criteria for noisy optimization.
		\newblock \emph{Structural and Multidisciplinary Optimization}, 48\penalty0
		(3):\penalty0 607--626, 2013{\natexlab{b}}.
		
		\bibitem[Picheny et~al.(2020)Picheny, Ginsbourger, and Roustant]{DiceOptim}
		V.~Picheny, D.~Ginsbourger, and O.~Roustant.
		\newblock \emph{DiceOptim: Kriging-Based Optimization for Computer
			Experiments}, 2020.
		\newblock URL \url{https://CRAN.R-project.org/package=DiceOptim}.
		\newblock R package version 2.0.1.
		
		\bibitem[Pirot et~al.(2015)Pirot, Straubhaar, and Renard]{Pirot2015}
		G.~Pirot, J.~Straubhaar, and P.~Renard.
		\newblock A pseudo genetic model of coarse braided-river deposits.
		\newblock \emph{Water Resources Research}, 51\penalty0 (12):\penalty0
		9595--9611, 2015.
		
		\bibitem[Pr{\'e}kopa(2013)]{Prekopa2013}
		A.~Pr{\'e}kopa.
		\newblock \emph{Stochastic programming}.
		\newblock Springer Science \& Business Media, 2013.
		
		\bibitem[Rasmussen and Williams(2006)]{RasmussenWilliams2006}
		C.~Rasmussen and C.~Williams.
		\newblock \emph{{G}aussian {P}rocesses for {M}achine {L}earning}.
		\newblock Adaptive {C}omputation and Machine Learning. MIT Press, Cambridge,
		MA, USA, 2006.
		
		\bibitem[Risk and Ludkovski(2018)]{Risk2018}
		J.~Risk and M.~Ludkovski.
		\newblock Sequential design and spatial modeling for portfolio tail risk
		measurement.
		\newblock \emph{SIAM Journal on Financial Mathematics}, 9\penalty0
		(4):\penalty0 1137--1174, 2018.
		
		\bibitem[Robbins and Monro(1951)]{Robbins1951}
		H.~Robbins and S.~Monro.
		\newblock A stochastic approximation method.
		\newblock \emph{The annals of mathematical statistics}, 22\penalty0
		(3):\penalty0 400--407, 1951.
		
		\bibitem[Rockafellar and Uryasev(2000)]{Rockafellar2000}
		R.~T. Rockafellar and S.~Uryasev.
		\newblock Optimization of conditional value-at-risk.
		\newblock \emph{Journal of risk}, 2\penalty0 (3):\penalty0 21--42, 2000.
		
		\bibitem[Rostek(2010)]{Rostek2010}
		M.~Rostek.
		\newblock Quantile maximization in decision theory.
		\newblock \emph{The Review of Economic Studies}, 77\penalty0 (1):\penalty0
		339--371, 2010.
		
		\bibitem[Ruszczy{\'n}ski and Shapiro(2003)]{Ruszczynski2003}
		A.~Ruszczy{\'n}ski and A.~Shapiro.
		\newblock Stochastic programming models.
		\newblock volume~10, pages 1--64. Elsevier, 2003.
		
		\bibitem[Srinivas et~al.(2010)Srinivas, Krause, Kakade, and
		Seeger]{Srinivas2009}
		N.~Srinivas, A.~Krause, S.~Kakade, and M.~Seeger.
		\newblock Gaussian {P}rocess optimization in the bandit setting: {N}o regret
		and experimental design.
		\newblock In \emph{Proceedings of the 27th International Conference on Machine
			Learning}, ICML'10, page 1015–1022, 2010.
		
		\bibitem[Talsk{\'a} et~al.(2018)Talsk{\'a}, Menafoglio, Machalov{\'a}, Hron,
		and Fi{\v{s}}erov{\'a}]{Talska2018}
		R.~Talsk{\'a}, A.~Menafoglio, J.~Machalov{\'a}, K.~Hron, and
		E.~Fi{\v{s}}erov{\'a}.
		\newblock Compositional regression with functional response.
		\newblock \emph{Computational Statistics \& Data Analysis}, 123:\penalty0
		66--85, 2018.
		
		\bibitem[Thompson(1933)]{Thompson1933}
		W.~R. Thompson.
		\newblock On the likelihood that one unknown probability exceeds another in
		view of the evidence of two samples.
		\newblock \emph{Biometrika}, 25\penalty0 (3-4):\penalty0 285--294, 1933.
		
		\bibitem[Tokdar(2007)]{TokdarTowards2007}
		S.~T. Tokdar.
		\newblock {T}owards a faster implementation of density estimation with
		{L}ogistic {G}aussian {P}rocess priors.
		\newblock \emph{Journal of Computational and Graphical Statistics}, 16\penalty0
		(3):\penalty0 633--655, 2007.
		
		\bibitem[Tokdar and Ghosh(2007)]{TokdarConsist2007}
		S.~T. Tokdar and J.~K. Ghosh.
		\newblock {P}osterior consistency of logistic {G}aussian process priors in
		density estimation.
		\newblock \emph{Journal of Statistical Planning and Inference}, 137\penalty0
		(1):\penalty0 34--42, 2007.
		
		\bibitem[Tokdar et~al.(2010)Tokdar, Zhu, and Ghosh]{tokdar2010}
		S.~T. Tokdar, Y.~M. Zhu, and J.~K. Ghosh.
		\newblock Bayesian density regression with logistic gaussian process and
		subspace projection.
		\newblock \emph{Bayesian Analysis}, 5\penalty0 (2):\penalty0 319--344, 2010.
		
		\bibitem[Torossian et~al.(2020)Torossian, Picheny, Faivre, and
		Garivier]{Torossian2020}
		L.~Torossian, V.~Picheny, R.~Faivre, and A.~Garivier.
		\newblock A review on quantile regression for stochastic computer experiments.
		\newblock \emph{Reliability Engineering \& System Safety}, 201\penalty0
		(C):\penalty0 106858, 2020.
		
		\bibitem[Zhu and Sudret(2020)]{zhu2020replication}
		X.~Zhu and B.~Sudret.
		\newblock Replication-based emulation of the response distribution of
		stochastic simulators using generalized lambda distributions.
		\newblock \emph{International Journal for Uncertainty Quantification},
		10\penalty0 (3):\penalty0 249--275, 2020.
		
	\end{thebibliography}
	
	}
\end{document}